# Magnetic Dirac semimetal state of (Mn,Ge)Bi$_2$Te$_4$

Alexander S. Frolov[1,2], Dmitry Yu. Usachov[1,3,4], Artem V. Tarasov[1,3], Alexander V. Fedorov[5], Kirill A. Bokai[1,3], Ilya Klimovskikh[1,3], Vasily S. Stolyarov[1,4], Anton I. Sergeev[1,6], Alexander N. Lavrov[7], Vladimir A. Golyashov[8,9], Oleg E. Tereshchenko[8,9], Giovanni Di Santo[10], Luca Petaccia[10], Oliver J. Clark[5], Jaime Sanchez-Barriga[5,12] and Lada V. Yashina[1,4,6]

[1] Center for Advanced Mesoscience and Nanotechnology, Moscow Institute of Physics and Technology, 9 Institutskiy Pereulok, National Research University, Dolgoprudny, Moscow Region 141700, Russia

[2] N.N. Semenov Federal Research Center for Chemical Physics, Kosygina Street 4, Moscow 119991, Russia

[3] St. Petersburg State University, 7/9 Universitetskaya nab., St. Petersburg 199034, Russia

[4] National University of Science and Technology MISIS, Moscow 119049, Russia

[5] Helmholtz-Zentrum Berlin für Materialien und Energie, Elektronenspeicherring BESSY II, Albert-Einstein-Str. 15, Berlin 12489, Germany

[6] Lomonosov Moscow State University, Leninskie Gory 1/3, Moscow 119991, Russia

[7] Nikolaev Institute of Inorganic Chemistry SB RAS, 3 Acad. Lavrentiev Ave, Novosibirsk, 630090, Russian Federation

[8] Rzhanov Institute of Semiconductor Physics, Siberian Branch, Russian Academy of Sciences, Novosibirsk, 630090, Russian Federation

[9] Synchrotron Radiation Facility SKIF, Boreskov Institute of Catalysis SB RAS, 630559 Kol'tsovo, Russia

[10] Elettra Sincrotrone Trieste, Strada Statale 14 km 163.5, 34149 Trieste, Italy

[11] IMDEA Nanoscience, 28049 Madrid, Spain

## Abstract

For quantum electronics, the possibility to finely tune the properties of magnetic topological insulators (TIs) is a key issue. We studied solid solutions between two isostructural $Z_2$ TIs, magnetic MnBi$_2$Te$_4$ and nonmagnetic GeBi$_2$Te$_4$, with $Z_2$ invariants of 1;000 and 1;001, respectively. For high-quality, large mixed crystals of Ge$_x$Mn$_{1-x}$Bi$_2$Te$_4$, we observed linear $x$-dependent magnetic properties, composition-independent pairwise exchange interactions along with an easy magnetization axis. The bulk band gap gradually decreases to zero for $x$ from 0 to 0.4, before reopening for $x>0.6$, evidencing topological phase transitions (TPTs) between topologically nontrivial phases and the semimetal state. The TPTs are driven purely by the variation of orbital contributions. By tracing the x-dependent 6$p$ contribution to the states near the fundamental gap, the effective spin-orbit coupling variation is extracted. As $x$ varies, the maximum of this contribution switches from the valence to the conduction band, thereby driving two TPTs. The gapless state observed at $x$=0.42 closely resembles a Dirac semimetal above the Néel temperature and shows a magnetic gap below, which is clearly visible in raw photoemission data. The observed behavior of the Ge$_x$Mn$_{1-x}$Bi$_2$Te$_4$ system thereby demonstrates an ability to precisely control topological and magnetic properties of TIs.

## 1. Introduction

Recently, magnetic topological insulators (MTIs) have become the focus of significant scientific interest. Among them, $Z_2$ topological insulators (TIs) with ordered magnetic moments of embedded magnetic atoms have attracted special attention. One particularly intriguing example is the case of systems with a surface-normal magnetic easy axis coexisting with a topologically nontrivial surface state. Such materials demonstrate the quantum anomalous Hall effect, which manifests itself as chiral edge conduction channels that can be manipulated by switching the polarization of magnetic domains.[1]

To date, the most intensely studied MTIs are MnBi$_2$Te$_4$ (the so-called 124 phase), a layered compound composed of septuple layers (SLs) with an idealized layer sequence of Te-Bi-Te-Mn-Te-Bi-Te, along with its homological compounds 147 and 1610 etc. For them, partial intermixing of the Mn and Bi positions is observed in the cation sublattice, yielding Mn$_{Bi}$ and Bi$_{Mn}$ antisite defects. MnBi$_2$Te$_4$ is an antiferromagnetic (AFM) TI with $T_N$=24.5K.[2] For such a system, gap opening is observed in the Dirac cone although its width and origin is still under debate.[2] The system can be switched to the QAH regime in the case of ultra-thin flakes.[3]

Potential practical applications of MnBi$_2$Te$_4$ are restricted by the availability of only small crystals due to preparative issues and limited thermodynamic stability.[4] The MnBi$_2$Te$_4$ lattice can be stabilized by mixing with stable isostructural compounds such as GeBi$_2$Te$_4$, SnBi$_2$Te$_4$, PbBi$_2$Te$_4$, which are believed to be $Z_2$ Tis.[5–7] Among them, GeBi$_2$Te$_4$ is ideally suited and meets all requirements for its crystal structure (cation coordination, bond lengths) and the spin-orbit coupling (SOC) strength (related to core charge effect). It is known that in the related system MnTe-GeTe, broad range of mutual solubility is observed for cubic Mn$_x$Ge$_{1-x}$Te, where Ge and Mn are also inside the Te octahedra.[8] One can expect similar behavior for the GeBi$_2$Te$_4$-MnBi$_2$Te$_4$ system which additionally holds the possibility to finely tune exchange interactions through dilution of the magnetic system. For instance, for Mn$_x$Ge$_{1-x}$Te, Mn substitution with Ge leads to a monotonic decrease of the Néel temperature in bulk crystals[9]. This behavior is further for the case of Mn$_x$Ge$_{1-x}$Te thin films, where electronic band structure is modified by strain.[10] Moreover Ge$_x$Mn$_{1-x}$Te films have a Curie temperature as high as 200 K for MnTe concentration of 50%.[11]

Regarding chemical bonding in the 124 compound, the main difference between MnBi$_2$Te$_4$ and GeBi$_2$Te$_4$ is the orbital contributions in the electronic bands. Ge atoms in GeBi$_2$Te$_4$ are bonded through 4$p$ electrons, while Mn atoms mainly contribute 4$s$ electrons. Therefore, the effective SOC of the valence electrons can be varied with changing composition, thereby potentially driving a SOC-mediated gap closing across the Ge$_x$Mn$_{1-x}$Bi$_2$Te$_4$ series and producing a 3D Dirac/Weyl semimetal (3D DSM/WSM) state for mixed crystals.[12] Both 3D DSM and WSM phases host a plenty of unique properties and phenomena, including high mobility of bulk carriers,[13] chiral anomaly and negative magnetoresistance,[14,15] Fermi arcs,[16] anomalous Hall effect,[17] charge-to-spin conversion,[18] spin Seebeck effect.[19] These effects have practical potential in topological electronics,[20] quantum computing, optoelectronics, spintronics and energy conversion.[21]

In this paper we report band structure and magnetic properties of Ge$_x$Mn$_{1-x}$Bi$_2$Te$_4$ crystals vs $x$, where the variation of $p$- and $s$- orbital contributions indirectly influences the Bi 6$p$ contribution to the bands near the fundamental band gap and induces a band inversion and a bulk gapless state corresponding to $x$=0.42. This semimetal state in turn demonstrates magnetic gap opening below Néel temperature. All observations were made for massive crystals of high structural perfection obtained by the original method developed in this work.

## 2. Results and discussion
### 2.1. Crystal growth and characterization

Initially, Ge$_x$Mn$_{1-x}$Bi$_2$Te$_4$ crystals were grown from melt using Bridgman method. Growth was performed from the melt of Ge$_x$Mn$_{1-x}$Bi$_2$Te$_4$ with $x$ taken equal to the targeted composition. However, massive bulk crystals of high quality were obtained only for $x$≥0.6. To grow crystals of lower $x$, we developed a new growth protocol with the source initially including two phases $L(Bi_2Te_3)+S(Ge_xMn_{1-x}Te)$. The solid phase was added as a compact tablet. Due to its lower density, the tablet floats over $Bi_2Te_3$ melt. After equilibration with the melt the tablet dissolves gradually during growth. Preliminary nucleation temperature was determined by differential thermal analysis. Growth was performed in evacuated quartz cylindrical ampoules with conic tips. The ampoule inner surface was preliminary graphitized. Each ampoule was placed into a furnace and equilibrated during a day at temperature 10°C higher than the nucleation temperature. Further, the ampoule was pulled in the temperature gradient with the rate of 0.2 mm/h. Growth conditions are summarized in Tables S1 and S2 of the Supporting Information. The result of this novel growth procedure produced good-quality, several-cm-sized crystals of whole composition range of this system.

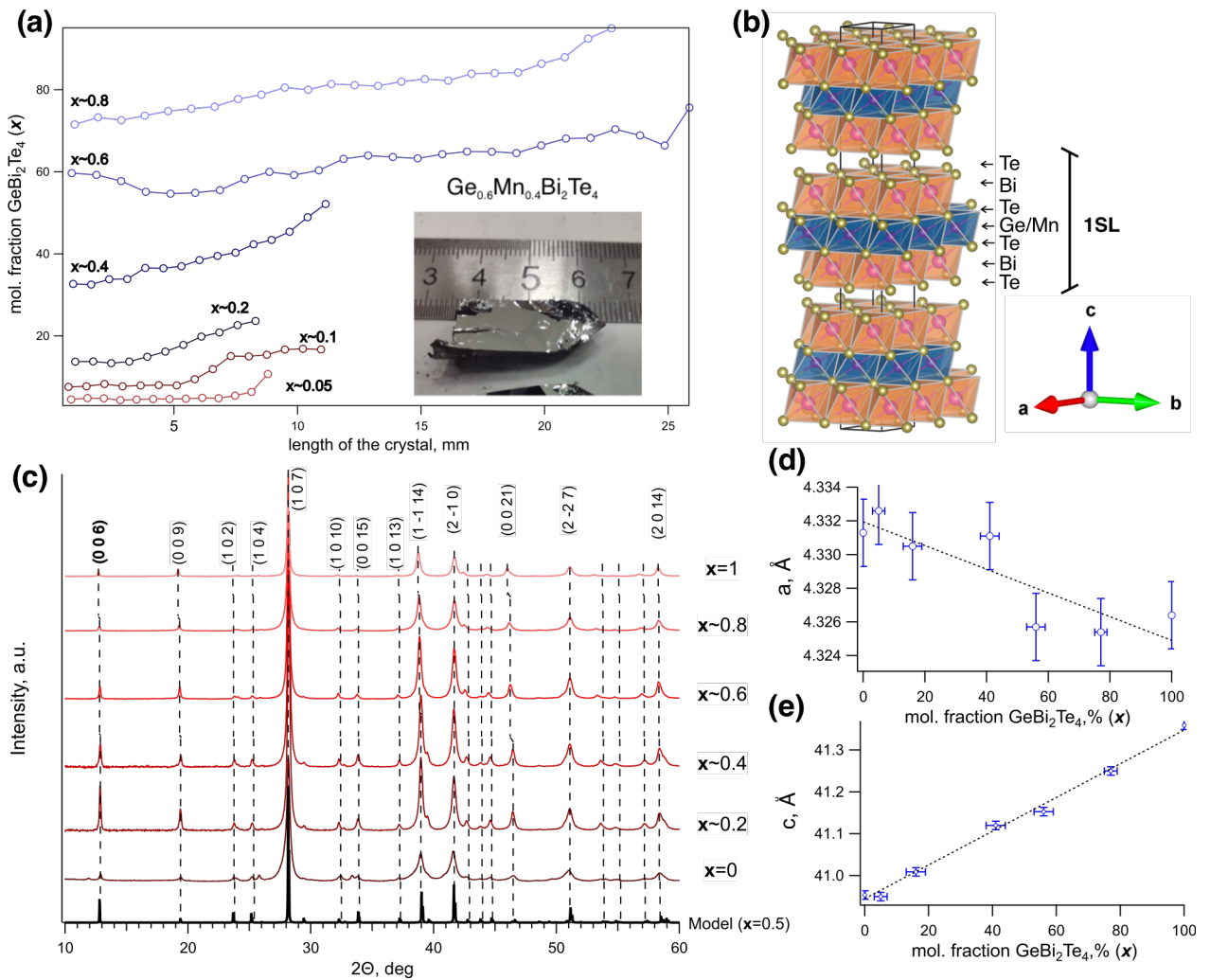

Figure 1. a) Crystal composition longitudinal variation and general view of a typical crystal (cleaved), b) crystal structure, c) X-ray diffractograms, d,e) lattice constants.

A typical crystal obtained is shown in Fig. 1a (inset). It should be noted that some crystals included in length several structural types of general formula $nGe_xMn_{1-x}Te*mBi_2Te_3$ grown one by one. Most of the experiments show the general tendency of $Y=(Ge+Mn)/(Ge+Mn+Bi) \approx n/(n+2m)$ to decrease during crystal growth. Fig. 1a exhibits $GeBi_2Te_4$ mole fraction along the longitudinal direction of 124 part for all crystals according to X-ray fluorescent spectroscopy. XRD data are collected in Fig. 1c. They correspond well to $R\underline{3}m$ structure (Fig. 1b). Both lattice constants, $a$ and $c$, vary slightly with composition (less than one percent) as shown in Fig. 1d,e.

## 2.2. Magnetic properties of Ge$_x$Mn$_{1-x}$Bi$_2$Te$_4$ crystals

In order to unveil the magnetic properties, the Mn charge state and its possible variation for different composition needs to be determined. To do this, x-ray photoemission spectroscopy (XPS) and absorption (NEXAFS) spectroscopy was employed. The corresponding Mn $2p$ spectra are presented in Fig. 2a and Fig. S1 of the Supporting Information. The absorption spectra show typical features I-IV which correspond to the transitions from $2p_{1/2}$ to split $3d$ levels, $e_g$ and $t_{2g}$. Photoemission spectra demonstrate a multiplet splitting typical of Mn$^{2+}$. No remarkable change of spectral shape is observed upon composition variation. All in all, both absorption and photoemission spectra and Mn $2p_{3/2}$ binding energies are similar to those of MnTe,[22] MnBi$_2$Te$_4$[4] and Mn$_{0.08}$Bi$_{1.92}$Se$_3$[23] evidencing a $3d^5$ configuration. More detailed analysis of photoemission spectra is complicated by complex background shape and limited intensity, especially for Ge enriched samples.

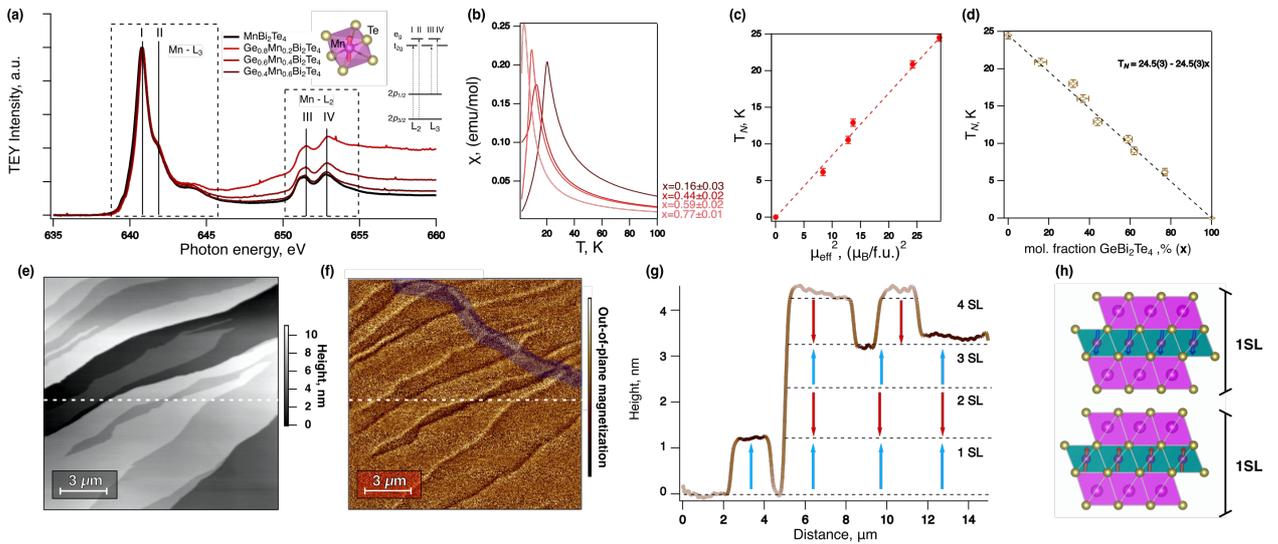

Figure 2. Mn L-edge absorption spectra (a) and magnetic properties of Ge$_x$Mn$_{1-x}$Bi$_2$Te$_4$ crystals: temperature dependence of magnetic susceptibility (b), Néel temperature dependence on effective magnetic momentum (c) and composition (d), Ge$_x$Mn$_{1-x}$Bi$_2$Te$_4$ ($x$=0.6) (111) surface topography (e) and magnetic contrast (f), surface profile and corresponding magnetic signal shown by curve color (g), crystal structure with indicated spin of Mn atoms (h).

Fig. 2b shows temperature dependences of magnetic susceptibility. Generally, all crystals are paramagnetic and demonstrate a transition to the AFM state with the easy magnetization axis, $c$, at the Néel temperature (additional data are available in Fig. S2 and Table S3 of the Supporting Information). Materials demonstrate AFM interlayer order and FM intralayer order as confirmed by magnetic force microscopy (MFM) results discussed below. Susceptibility curves for all compositions have a similar shape, and even the most dilute magnetic crystal of $x$=0.77 ± 0.01 demonstrates behavior similar to that of pure MnBi$_2$Te$_4$. Upon dilution of the magnetic sublattice, the Néel temperature drops linearly from $T_N$≈20 K for $x$=0.16±0.03 to $T_N$≈6 K for $x$=0.77±0.01 (see Fig. 2d). Besides, Néel temperature is also a linear function of squared effective magnetic moment.

Consequently, the effective exchange integral, in turn, is a linear function of dilution $x$; and a pairwise exchange coupling does not depend on $x$. This is true even for the most dilute crystal showing AFM properties. These observations demonstrate that the easy magnetization axis does not depend on composition, and magnetic dilution affects neither the choice of magnetically active sublattices nor the pairwise exchange interaction of ions.

To verify our supposition on magnetic structure, namely on interlayer and intralayer interactions, we studied a Ge$_x$Mn$_{1-x}$Bi$_2$Te$_4$ crystal with $x$=0.6 by MFM. Figures 2e and 2f show topography with terraces of SL height and the corresponding magnetic contrast below Néel temperature. One can see that the magnetic contrast appears at the steps of SL height (or odd number of SLs). There are also additional antiphase boundaries which are not related to the steps, an example of which is indicated with a violet line in Fig. 2f. Correspondence between the magnetic contrast and the step height is in line with AFM order between ferromagnetically ordered SLs as schematically shown in Fig. 2h. All in all, the behavior of Ge$_x$Mn$_{1-x}$Bi$_2$Te$_4$ enables precise control and fine tuning of magnetic properties of topological insulator by $x$ variation.

### 2.3. Electronic band structure of Ge$_x$Mn$_{1-x}$Bi$_2$Te$_4$

Now we focus on the effect of Mn substitution with Ge on the electronic band structure, and its evolution for mixed crystals of Ge$_x$Mn$_{1-x}$Bi$_2$Te$_4$ with $x$ ranging from 0 to 1. Figure 3 shows angle-resolved photoemission spectroscopy (ARPES) results obtained with a photon energy of 21.2 eV at which the photoemission (PE) cross section for the topological surface states (TSS) of MnBi$_2$Te$_4$ is rather low,[24] and PE signal from the bulk bands dominates in the spectra. The upper row in Fig. 3 presents band dispersions along the K-Γ-K direction of the surface Brillouin zone (BZ). In general, the ARPES data for all $x$ values exhibit rather similar sets of bands. The bulk valence band (VB) has a nearly cone-like shape. Its constant-energy contours are represented by hexagonally warped circles with a decrease of warping close to the top of the band. The bulk character of this band is evidenced by its $k_z$ dispersion.[24,25] In the range of $x$=0.2-0.8, this band demonstrates a monotonic shift towards the Fermi level that can be assigned to gradual depopulation of the conduction band states. At $x$=1, corresponding to GeBi$_2$Te$_4$, the material again becomes $n$-doped most probably due to another type of dominant point defects. Generally, GeBi$_2$Te$_4$ crystals can have variable carrier concentration depending on nonstoichiometry:[26] Te excess provides a lower carrier concentration as shown in Fig. S4 where ARPES data are presented for crystals with different composition.

The bulk conduction band (CB) for Ge$_x$Mn$_{1-x}$Bi$_2$Te$_4$ for low $x$ is of a nearly conical shape and can be regarded as a counterpart of the VB (see Fig. 3). It should be noted that the gap between these states observed with ARPES strongly depends not only on the crystal composition but on the photon energy as well, and may be larger than the fundamental gap because of the $k_z$ dispersion of VB and CB states. In the range of $x$=0-0.17 the observed gap is about 0.12 eV, while at $x$=0.38 the gap shrinks. Due to the finite width of the PE spectral features, it is therefore unclear whether the gap becomes fully closed or not, but an upper limit of 0.05 eV can be placed with certainty. Upon further increase of $x$ to 0.65 the ARPES data indicate no visible opening of the gap, while at *x*=0.81 the VB apex moves above the Fermi level and the gap becomes inaccessible for PE analysis.

Near the Fermi level, one can also see the band labeled as SS, which is barely visible at *x*=0, but becomes more pronounced with increasing Ge content. As it was shown for MnBi$_2$Te$_4$, this band is formed by surface states which demonstrate Rashba-type splitting[27] combined with magnetic exchange splitting[28]. Density functional theory (DFT) calculations predicted that these surface states should remain in the case of Mn$_{0.5}$Ge$_{0.5}$Bi$_2$Te$_4$[29] in full consistency with our ARPES data in Fig. 3. Upon the increase of *x* the SS band appears closer to VB and reaches it at *x*=0.65.

Our ARPES data for pure GeBi$_2$Te$_4$ (*x*=1) demonstrate a characteristic topological surface state (TSS) that appears in agreement with the previous studies of this compound.[30] The conduction band here is formed by the bands S1 and S2, which were assigned as bulk states on the basis of their photon energy dependence.[30] Their W-like form at the Γ-point indicates an indirect character of the fundamental band gap that was estimated to be 0.12 eV wide.[30] In our data in Fig. 3, the apparent gap width (at *hv*=21.2 eV) is about 0.17 eV. One can see that the TSS touches the VB, so that its Dirac point appears below the VB maximum. Gradual energy shift of SS bands towards the VB upon the increase of Ge content suggests that the Rashba-type SS band, which is trivial at *x*=0,[27] evolves into the nontrivial TSS of GeBi$_2$Te$_4$. Taking into account that this transformation is accompanied by shrinkage and reopening of the bulk gap, we suggest that topological phase transitions (TPTs) occur at *x* value of 0.5±0.1.

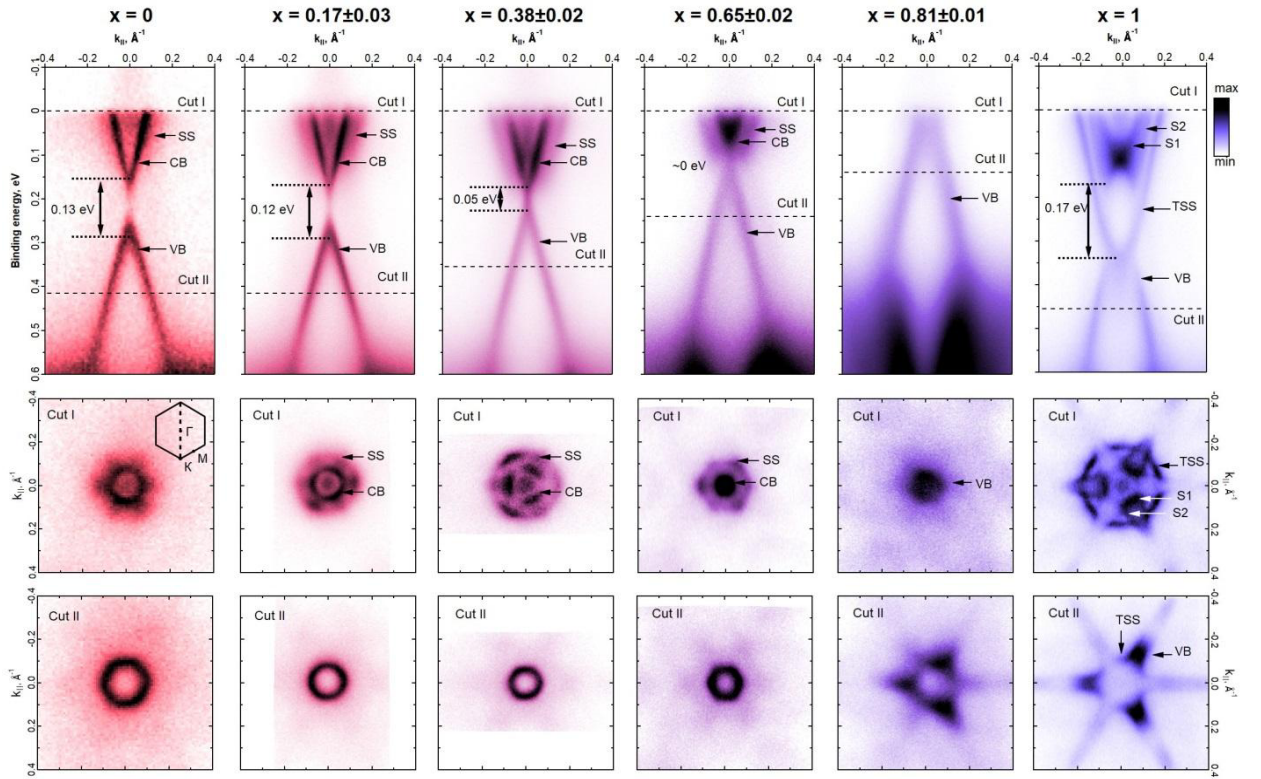

Figure 3. Electronic structure of topological insulators Ge$_x$Mn$_{1-x}$Bi$_2$Te$_4$ measured by ARPES as a function of Ge content *x*. The upper row shows band dispersions measured along the high-symmetry K-Γ-K direction. The middle row displays Fermi-surface maps. The bottom row contains constant-energy maps plotted for different energies marked as "cut II". The photon energy was 21.2 eV (He I radiation, 15 K).

Now we consider the photon energy dependencies of the PE spectra. This type of measurements gives information on the dispersion of electronic states in the direction perpendicular to the sample surface (Γ-Z direction in the Brillouin zone), thus allowing to distinguish bulk states from surface states which have no dispersion along Γ-Z ($k_z$ direction). The analysis is based on the assumption that the electron is excited into the free-electron-like state inside the crystal with constant inner potential. $V_0$. In this case, the photoelectron energy is directly connected with its momentum in the ground state. Figure 4 shows the dependencies of the normal-emission spectra on the photon energy for different samples in the paramagnetic (PM) phase. In such a geometry, the in-plane components of momentum are fixed to zero, while the out-of-plane component $k_z$ is defined by the photon energy as

$$\hbar^2 k_z^2/2 = h\nu + V_0 - \varphi - E_B, \qquad (1)$$

where $\varphi$ is the work function and $E_B$ is the binding energy. The experimentally observed periodicities in our $k_z$ dispersions are well described under the assumption that $V_0 = \varphi + E_B$. The respective relation between the photon energies and high-symmetry points is indicated above the bottom scales in Fig. 4. In our notations, $Γ_0$ corresponds to zero $k$. The $k_z$ dependence for MnBi$_2$Te$_4$ presented in Ref. [25] indicates that the gap between the VB and CB varies between 0.13 eV at the Γ point and 0.20 eV at Z. Our data for $x$=0.16 (left panel of Fig. 4) show qualitatively similar gap variations, although the bottom of the CB is not well distinguished. The photon energy dependence changes significantly at $x$=0.42. At this concentration the bottom of the CB becomes well resolved and its $k_z$ dependence closely resembles that of the top of the VB. Analysis of individual ARPES maps taken at different $h\nu$ indicates that the CB minimum, which is observed at the Z point, is located at nearly the same energy as the VB maximum observed in the Γ point. Thus, at $x$=0.42 the system turns into a semimetal with a zero indirect bandgap in the PM phase. At $x$=0.56 the amplitude of the $k_z$ dispersion increases for the VB, while we do not detect any significant dispersion for the CB. Upon further increase of Ge content to $x$=0.76 the VB dispersion does not change much in its shape, although it is shifted to the Fermi level; the system becomes p-doped and the CB appears completely unoccupied. Finally, at $x$=1 the system becomes n-doped again. In its $k_z$ dispersion we recognize the dispersive VB and non-dispersive TSS marked with the dotted line in the right panel of Fig. 4. The non-dispersive character of the TSS confirms its localization at the surface.

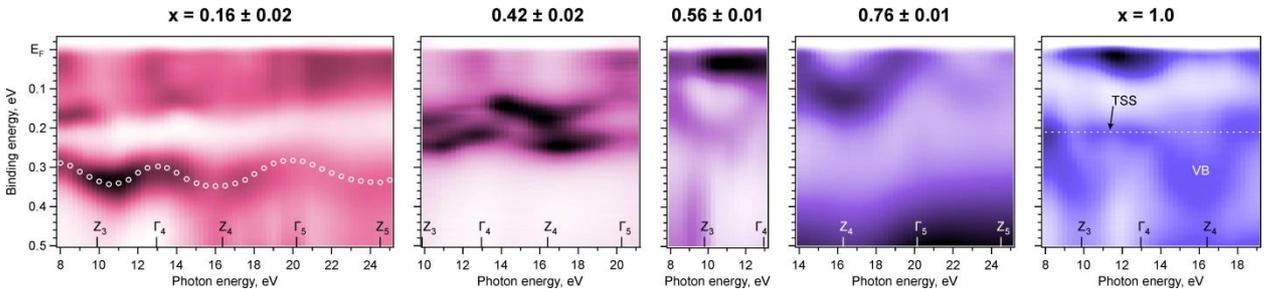

Figure 4. Dependence of the normal-emission PE spectra on the photon energy for Ge$_x$Mn$_{1-x}$Bi$_2$Te$_4$ with different Ge concentrations in the PM phase (at 20 K).

All in all, for mixed crystals of Ge$_x$Mn$_{1-x}$Bi$_2$Te$_4$, upon composition variation we observed experimentally a bulk band gap closure around $x$=0.4 and a further opening at $x$>0.65 that may indicate TPTs between two topologically nontrivial materials. It is remarkable that this transition occurs neither due to strong SOC variation nor structural changes, since both system components ($MnBi_2Te_4$ and $GeBi_2Te_4$) are rather similar in these properties. To shed light on the underling mechanism, we performed a theoretical study of the electronic structure. First, we used a coherent potential approximation (CPA) in frames of the Korringa–Kohn–Rostoker (KKR) Green's function method, which enables adequate modeling of mixed crystals by taking into account disorder induced by intermixing of Mn and Ge. Within this approach crystal composition can vary smoothly without construction of large supercells. To trace purely compositional effects in our calculations atomic positions for mixed crystals were fixed as they are known for $MnBi_2Te_4$.[2] The results are collected in Fig. 5 where the KKR-calculated bulk band structure of $MnBi_2Te_4$ along the K-Γ-Z direction in the BZ and its variations upon gradual substitution of magnetic Mn atoms with non-magnetic Ge atoms are shown, the spectral functions being with a minimal half-width of about 15 meV. According to these calculations, the bulk gap gradually decreases to zero for $x$ variation from 0 to 0.4. For $x>0.6$ a gap opens and it increases while $x$ approaches unity. The dependencies of the gap width in the Γ and Z points on crystal composition are exhibited in Figs. 5(i-j). These calculations reproduce our ARPES results despite both a slight difference in structure of counterparts and the presence of Bi/Ge (and Bi/Mn) antisite defects which were disregarded theoretically. The gapless states at $x = 0.4$—0.6 may pinpoint TPTs from a topologically nontrivial to a trivial phase or semimetal state with a zero bulk gap and further transition to TI above $x =0.6$.

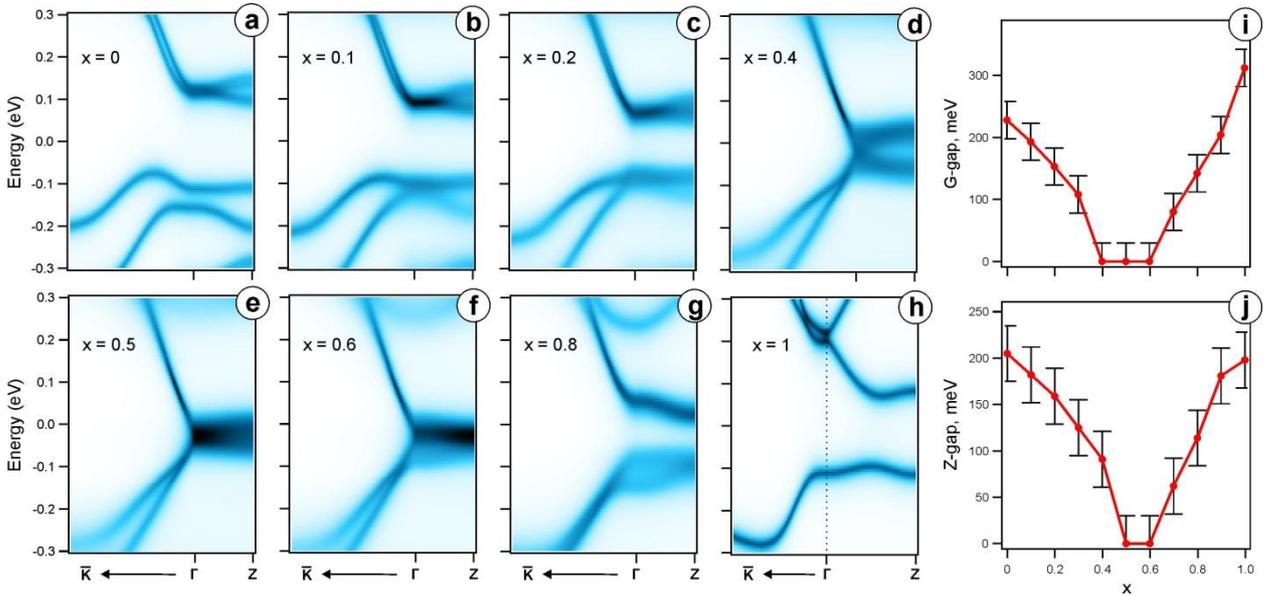

Figure 5. (a-h) Bulk band structure of $Ge_xMn_{1-x}Bi_2Te_4$ in the AFM phase (except for nonmagnetic system with $x=1$) calculated with the KKR approach. (i-j) Gap width in Γ (i) and Z (j) points of the BZ.

The underlying mechanism of TPTs may be related to the difference in chemical bonding of Mn and Ge atoms in 124 compounds. This difference is due to the fact that Mn atoms form bonds *via* 4s electrons while Ge uses 3p electrons. To evaluate orbital contributions into bands near Fermi level, we performed additional DFT modelling in this part of the study. This approach does not allow simulation of randomly intermixed Ge and Mn atoms, for this reason we considered 2x2 supercells for certain $x$ values of 0.25, 0.5 and 0.75. In these calculations the atomic positions were also fixed being equal to those of pristine MnBi$_2$Te$_4$, like in KKR calculations discussed above, with single exception - vdW gap between SLs in MnBi$_2$Te$_4$ crystal was increased by 0.16 Å for better correspondence with ARPES data on $k_z$ dispersion. The DFT calculation results are illustrated in Fig. 6. Similarly to the KKR data discussed above, here bulk band gap becomes closed at the Γ point at about $x$=0.5 upon Mn substitution with Ge, followed by a gap reopening in Γ when $x$ approaches unity. However, in the DFT calculation for $x$=1 the gap appears to be almost closed between the Γ and Z points in contrast to the KKR and experimental results. One of the possible reasons is the strong sensitivity of the DFT calculated electronic structure of GeBi$_2$Te$_4$ to computational geometry. Indeed, the modeling of GeBi$_2$Te$_4$ with its own experimental crystal structure[31] results in the gapped state as it is clearly seen in Fig. 6(f). Moreover, to ensure that gap closure at $x$=0.5 is not the result of the artificially fixed structural parameters, we performed additional DFT calculations using the averaged cell parameters of MnBi$_2$Te$_4$ and GeBi$_2$Te$_4$. These calculations verified that the gap remains closed in this case (see Fig. S4 of the Supporting Information).

It is worth noting that in the experimental ARPES data, the transition from the PM to the AFM state leads to minor changes of the spectral features, while in the calculation the formal period of the unit cell changes from 1 SL to 2 SL. This leads to doubling of all bands due to their folding. For this reason, in Fig. 6(h) the VB of nonmagnetic GeBi$_2$Te$_4$ is represented with a single branch, while for the AFM systems in Figs. 6a-g one can see two branches. To allow direct comparison with the ARPES data, we showed in Figs. 6a-d the bands unfolded to the unit cell of 1 SL height. In such presentation one can clearly see where the VB undergoes exchange splitting upon AFM ordering. Two examples of such exchange splitting are marked as Δ$_2$ and Δ$_3$ in Fig. 6b. They were observed in the ARPES data of MnBi$_2$Te$_4$,[25,32] although our DFT calculations overestimate their experimental values (Δ$_2$≈25 meV and Δ$_3$≈45 meV at $x$=0 [32]).

To disentangle the question of the gap closing and reopening upon variation of the Ge content, we trace the changes of the band orbital composition represented by the weight of Bi $p$-orbitals in Figs. 6(g-k). Looking at the electronic states which form the bulk gap, it becomes evident that as $x$ passes through the value of 0.5 the dominance of Bi-$p$ at the VB maximum near the Γ point is replaced by its dominance at the CB minimum. In other words, there is an inversion of the Bi $p$-orbitals contribution, indicating a possible transition from the TI phase into a semimetal state. Further increase of Ge concentration up to 100 % leads to a new inversion of the Bi $p$-orbitals contribution, but near the Z point, which indicates return of the system into the TI state. Summing all mentioned above, it possible to suggest two possible phase transitions in Ge$_x$Mn$_{1-x}$Bi$_2$Te$_4$ as value of $x$ increases: (i) from TI to semimetal with increasing $x$ value above 0.5 and (ii) from semimetal to TI when $x$ approaches unity.

The analysis of changes in the Bi-$p$ contributions in the bulk gap edges at the Γ and Z points deserves special attention. It is evident from Figs. 6(l,m) that the inversion of the Bi-$p$ contribution, occurring at some points of the intervals 0.5 < x < 0.75 at Γ point and 0.75 < x < 1 at the Z point, is accompanied by a general decrease in the Bi-$p$ contribution at the Γ point and its increase at the Z point. Since Bi is the main source of SOC in the system it can be assumed that the above mentioned TPTs should be related to the modulation of the effective SOC strength at the Γ and Z points.

It should be noted that TPTs cannot be revealed for materials of the same topological class. To clarify the topological properties of GeBi$_2$Te$_4$ and MnBi$_2$Te$_4$ we used a Wilson Loop technique and calculated Wannier Charge Centers (WCCs).[33] The results show different topological classes, namely (1;000) for AFM MnBi$_2$Te$_4$ and (1;001) for GeBi$_2$Te$_4$ (Fig S5,6.). This is in accordance with previous calculations.[34] This explains the observed behavior in terms of topological band theory. In detail, upon *x* variation two situations are possible: i) transition to trivial state (0;000), ii) transition TI state (0;001). Scenario i) is realized when the gap is first closed in Γ point and further in Z point. Contrary, case ii) is implemented when the gap is closed in Z point first. Our KKR calculations firs better to the case i). Unfortunately, due data sensitivity to tiny variation of structural parameters this conclusion is not rigorous.

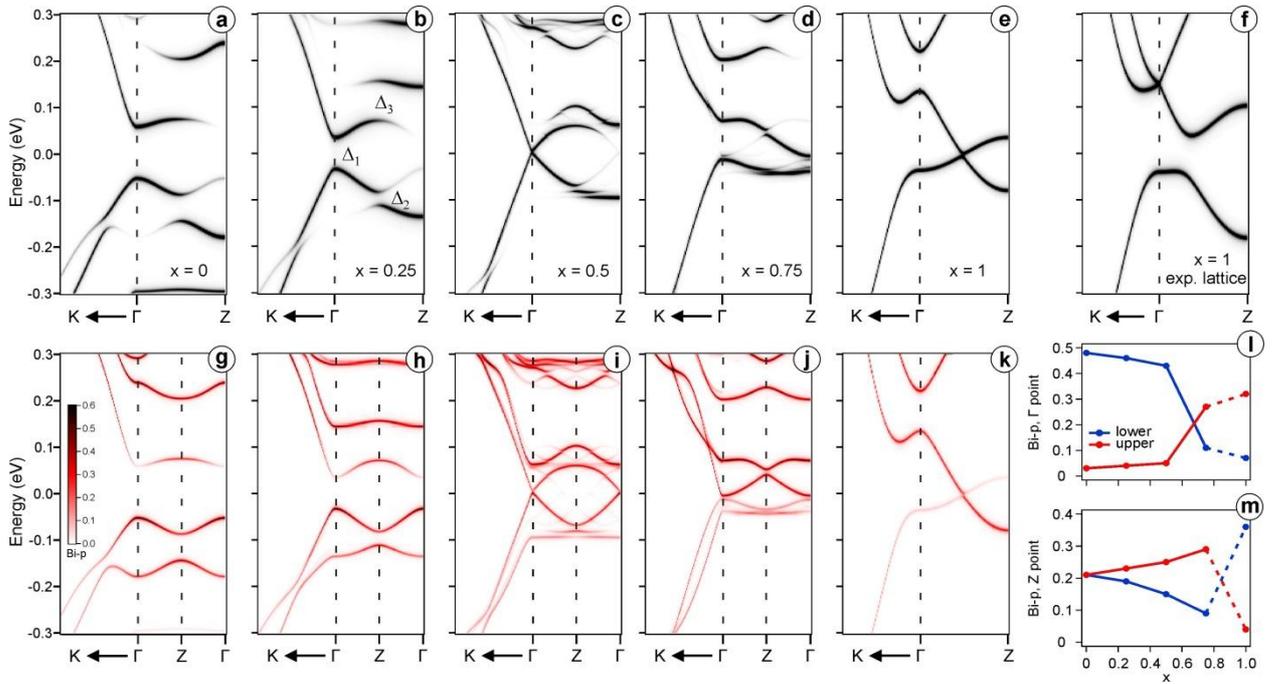

Figure 6. Electronic structure of Ge$_x$Mn$_{1-x}$Bi$_2$Te$_4$ supercells in the AFM phase calculated with DFT. The gray color scale in (a-e) indicates the total band weight unfolded to a primitive 1x1 unit cell with the height of 1 SL. The color scale in (g-k) illustrates the weight of Bi *p*-orbitals; in (h-j) the weight is unfolded to a 1x1 cell with the height of 2 SL; in (g) and (k) the weight is given without unfolding for the unit cells with the heights of 2 SL and 1 SL, respectively. In (f) the bands of GeBi$_2$Te$_4$ were calculated for its own lattice parameters in contrast to (a-e), where the lattice parameters of MnBi$_2$Te$_4$ were used. (l,m) The weight of Bi *p*-orbitals in the Γ and Z points.

Next, we discuss the nature of the gapless state of Ge$_x$Mn$_{1-x}$Bi$_2$Te$_4$. For this let us now consider in detail the band structure of the system with *x*=0.42 when Ge$_x$Mn$_{1-x}$Bi$_2$Te$_4$ becomes a semimetal. Our measurements of magnetic properties discussed above indicate that this crystal undergoes a transition from the PM to the AFM state at 13 K. Figure 7a shows a set of high-resolution temperature-dependent ARPES spectra that indicate opening of a tiny bandgap around the Γ point below Néel temperature, while at 25 K no traces of the gap are present in the data. This is clearly visible in the energy distribution curves (EDCs) in Fig. 7c that clearly demonstrate the magnetic nature of the gap. The gap appears to be M-shaped and its width is about 20 meV. It is essential to note that the momentum distribution curves (MDCs) of the VB given in Fig. 7b cannot be described with a single band, but they show a multi-component structure that can be approximated by at least two bands (red and blue lines) differing in energy by ~30 meV. Such splitting is also observed for the CB, although this is not illustrated. We did not find any significant difference between the splitting values in the PM and AFM phases; therefore, it is not related to magnetism. The reason of such multi-component structure is related to the fact that the $k_z$ component of (quasi)momentum is not conserved in photoemission, therefore the ARPES data contain contributions from the electronic states with all possible $k_z$, although with different intensities. Of course, the most intense contribution is expected from the $k_z$ value defined by Eq. 1. Nevertheless, superposition of different $k_z$ leads to an apparent splitting with its value nearly equal to the amplitude of the $k_z$ dispersion.

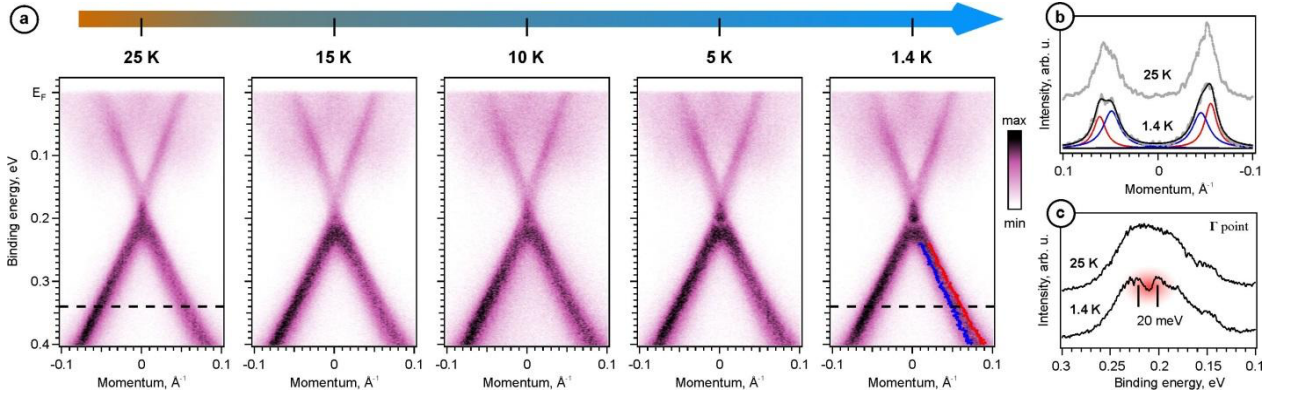

Figure 7. (a) Temperature-dependent ARPES data for $x=0.42\pm0.02$ along the M-Γ-M direction at $hv$=12 eV. (b) MDCs of the VB along the dashed lines in the panel (a). (c) Comparison of EDCs in the Γ point at different temperatures.

Thorough analysis of the ARPES spectra taken at various photon energies and temperatures allowed us to propose the electronic structure of the semimetal phase at $x=0.42$ schematically shown in Fig. 8, where the bands are presented along the in-plane direction at high-symmetry points and out-of-plane direction as a function of $k_z$. As we already noted, in the PM phase the VB and CB disperse synchronously and form a nearly zero indirect gap, whereas the local direct gap $\Delta_0$ is present at any $k_z$ and its width is about 40 meV. Upon transition to the AFM phase the lattice period in the c (or z) direction gets doubled due to opposite directions of magnetic moments in neighbor SLs. This leads to shrinkage of the BZ and folding of bands. As a result, the bands originally located in the Z point are translated to the Γ point where the CB and VB may have touched each other. However, instead of touching the bands demonstrate avoided-crossing behavior and the 20 meV gap marked as $\Delta_1$ opens between them. The material, thus, becomes a semiconductor in the AFM phase. At the same time, the in-plane dispersion of the VB in the Γ point gains M-like shape. This seems to be a result of hybridization between the VB and CB states. Hybridization is evident from the photoemission intensity distribution. In particular, one can see that the VB is more intense than the CB (at $hv$=12 eV) and the high intensity of the VB spreads slightly above the gap $\Delta_1$. This is a clear indication of admixture of the orbital character of the VB to the bottom of the CB. In our scheme, we also admitted possible splitting of the VB and CB in the $Z_{2SL}$ points at the boundary of the AFM BZ (see Fig. 6); these gaps are marked as $\Delta_2$ and $\Delta_3$, respectively. The splitting is well seen in the AFM phase of MnBi$_2$Te$_4$ at the photon energy of 6.3 eV.[25,32] Since it has a magnetic nature, it is expected to be notably smaller in the Ge$_x$Mn$_{1-x}$Bi$_2$Te$_4$ crystals with large $x$. This is a probable reason why we do not resolve $\Delta_2$ and $\Delta_3$ in our ARPES data.

Temperature behavior of the experimentally observed $\Delta_1$ gap size is illustrated in Fig. S6 Generally, it obeys a power law $E_g(T) = E_{g0}\cdot(1-T/T_N)^\beta$. According to Ising model, the β value is expected to be β=1/8[35] and β≈0.3[36] in the case of 2D and 3D systems respectively. In our case β=1/8 corresponds better to the experimental observations. Unfortunately, the experimental statistics is insufficient to make any solid conclusions on this matter.

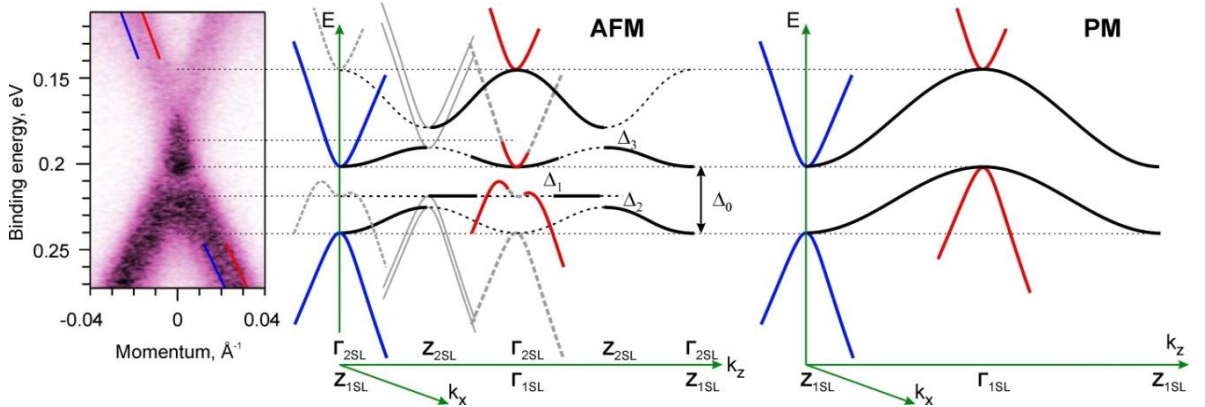

Figure 8. Scheme of the bulk electronic structure of Ge$_x$Mn$_{1-x}$Bi$_2$Te$_4$ at $x$=0.42 in the PM and AFM phases based on the ARPES data.

Taking into account our understanding of bulk band structure of Ge$_x$Mn$_{1-x}$Bi$_2$Te$_4$ with $x$ up to 0.42, a next question arises whether the TSS appear in this system. To clarify this point, we explored the systems with $x$=0.16 and 0.42 with ARPES at $hv$=6.3 eV, which is known to be suitable for efficient observation of TSS in MnBi$_2$Te$_4$. The data are displayed in Fig. 9. One can see that for $x$=0.16 a quite intense TSS is present. The temperature of 15 K reached in this experiment is lower than the Néel temperature of about 20 K for this sample, but we did not observe clear traces of a magnetic gap opening in the TSS upon changing the temperature between 30 K and 15 K. For the sample with $x$=0.42 we clearly see again that these bands have a multicomponent structure due to integration over $k_z$-dimension. We also see where the bands overlay forming a diamond-shaped region from 0.24 eV to 0.19 eV (best seen in second derivative). As we have shown in Fig. 8, the VB and CB overlay at 0.20±0.01 eV indicating that some extra states contribute to the higher BE part of the diamond-shaped region. Since the direct gap of at least $\Delta_0$=40 meV is present in the sample in the PM phase, it is natural to suppose that the extra contribution is due to the presence of TSS within the gap. Although the Rashba-type SS is clearly visible, the signal from TSS is weak in the system with $x$=0.42. We suppose the following reason for this. As the fundamental gap shrinks upon the increase of Ge content, the spatial localization of the TSS shifts more and more inside the bulk. When the gap becomes as small as several dozens of meV, the TSS becomes localized in a rather extended region below the surface, so it becomes almost indistinguishable from the bulk states.

Summing up, we have revealed and explained the general phenomenon of TPTs, also observed also in similar systems, where Mn is substituted with group XIV elements, Sn[37] and Pb.[38] In these systems, a decrease of the bulk gap down to zero was observed in some ranges of the $x$ values. The authors of these works also suggested possible TPTs in these systems. In contract to Sn and Pb, we consider an ideal model system in which effect of geometry and atomic number on SOC is switched off, that enables to trace delicate SOC variation due to the change of the Bi 6$p$ orbital contribution in the electronic bands near the fundamental gap. This change is caused indirectly by the substitution of the Mn 4$s$ contribution with Ge 4$p$ contribution to chemical bonds.

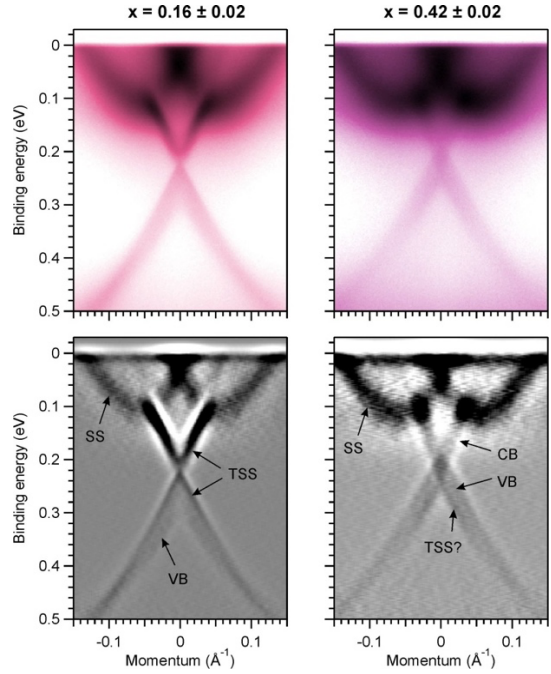

Figure 9. ARPES maps (upper row) obtained with laser radiation of 6.3 eV and corresponding second derivatives of the PE intensity (lower row).

## 3. Conclusions

We studied solid solutions between two $Z_2$ TIs, magnetic MnBi$_2$Te$_4$ ($Z_2$ invariants are 1;000) and nonmagnetic GeBi$_2$Te$_4$ ($Z_2$ invariants are 1;001), which differs essentially neither in structural parameters nor in atomic core charges affecting SOC strength of the valent electrons. We have considered an ideal model system where geometry and atomic number effects are switched off, that enables to trace chemical effects solely.

We observed a linear *x*-dependence of magnetic properties (Néel temperature), which points to composition-independent pairwise exchange interactions. Moreover, the easy axis does not depend on composition, and magnetic dilution affects neither the choice of magnetically active sublattices nor the pairwise exchange interaction of ions. MFM confirms antiferromagnetic order between ferromagnetically ordered SLs. The behavior of Ge$_x$Mn$_{1-x}$Bi$_2$Te$_4$ enables precise control and fine tuning of magnetic properties of topological insulator by *x* variation.

In the electronic band structure probed by ARPES and theoretically the bulk band gap gradually decreases to zero for *x* variation from 0 to 0.4. For *x*>0.6 a gap opens and it increases while *x* approaches unity. This pinpoints two TPTs from a topologically nontrivial phase to a semimetal state, that is possibly a Dirac semimetal state, and back to TI phase. The TPTs are driven solely by the variation of orbital contributions to chemical bond (3p for Ge and 4s for Mn) and, in turn, indirectly to the states near the fundamental gap. Since Bi is the main source of SOC in the system it can be assumed that the above mentioned TPTs should be related to the modulation of the effective SOC strength. Using the Bi 6*p* contribution to the states near the fundamental gap, we traced effective SOC variation that causes band inversions at the Γ and Z points at two different values of *x*, thus producing two TPTs upon *x* variation.

Gapless state observed at *x*=0.42 closely resembles a Dirac semimetal state. It shows a magnetic gap below Néel temperature which is clearly visible in raw ARPES data. We propose the detailed electronic structure of the semimetal phase. As the fundamental gap shrinks upon the increase of Ge content, the spatial localization of the TSS shifts more and more inside the bulk. When the gap becomes as small as several dozens of meV, the TSS becomes localized in a rather extended region below the surface, so it becomes almost indistinguishable from the bulk states.

## 4. Methods

All crystals were grown from melt using modified Bridgeman method in the four-zone vertical furnace from the mixture of binary compounds $Bi_2Te_3$ and $Ge_{1-x}Mn_xTe$. Raw materials were synthesized by prolonged annealing of stochiometric mixtures of the elements. The obtained crystals were characterized by XRF and additionally by powder XRD. The XRF measurements were performed using a Bruker Mistral-M1 micro-focused system equipped with an XFlash 30mm$^2$ detector, the concentrations were determined using XSpect software by determining the area of individual peaks using the external standard model. 4–7 measurements were taken for each crystal. For XRD measurements the samples were powdered. XRD was performed using a Rigaku Smartlab SE diffractometer equipped with a PSD Pixel3D detector in the 2θ range of 10–80° with a step of 0.02° at room temperature. The data obtained were treated within Jana 2006 software package.

Magnetic susceptibility of the $Ge_xMn_{1-x}Bi_2Te_4$ crystals was studied in the temperature range of 1.77–300 K using a Quantum Design MPMS-XL and Quantum Design MPMS SQUID VSM magnetometer. Measurements were performed under magnetic fields applied along or transverse to the crystallographic *c* axis. To determine the effective magnetic moment $\mu_{eff}$, the paramagnetic component of the magnetic susceptibility, $\chi_p(T)$, was analyzed using the Curie-Weiss expression $\chi_p(T) = N_a\mu_{eff}^2/3k_B(T-\vartheta)$. A part of measurements was performed at the Center of Diagnostic of Functional Materials for Medicine, Pharmacology, and Nanoelectronics of the Research Park of St. Petersburg State University.

Magnetic force microscopy measurements were performed using AttoDry 1000 MFM system at the temperature T=4.1K using Bruker MESP-LN-V2 probes covered by CoCr coating.

ARPES data were recorded using several facilities. Laboratory-based measurements were made using a SPECS GmbH ProvenX-ARPES system located in ISP SB RAS equipped with ASTRAIOS 190 electron energy analyzer with 2D-CMOS electron detector and a non-monochromated He Iα light source with hv=21.22 eV. Photon energy dependence of normal-emission PE spectra was measured at BaDElPh[39] beamline of the Elettra synchrotron in Trieste. All measurements were carried out at T = 12K using p-polarized photons. Temperature-dependent ARPES measurements were carried out at UE112-PGM2a undulator using ARPES 1^3 endstation of the BESSY II synchrotron in Berlin. Laser-based data was collected using laser ARPES endstation equipped with VG Scienta SES R4000 hemispherical analyzer at Hiroshima Synchrotron Radiation Center (HiSOR). Spin-resolved ARPES measurements were performed with linearly polarized light at the U125-2-PGM RGBL Undulator using the RGBL-2 endstation. The endstation was equipped with hemispherical Scienta R4000 analyzer. The RGBL-2 endstation is equipped with a combined detector that comprises a 2D channelplate for ARPES and a Mott-type spin-detector,[40] which was operated at 25 kV. The energy and angular resolutions of spin-resolved ARPES measurements were 45 meV and 0.75°, respectively. XPS and NEXAFS measurements were performed at the Russian– German dipole beamline (RGBL) of the synchrotron light source BESSY-II at Helmholtz-Zentrum Berlin. The photoemission spectra were acquired with a hemispherical SPECS Phoibos 150 electron energy analyzer at normal emission geometry. NEXAFS spectra were recorded in the total-electron yield mode. The photon energy was calibrated using second-order reflection. All core-level spectra were fitted by Gaussian/Lorentzian convolution functions with simultaneous optimization of the background parameters using the UNIFIT 2014 software.[41] The background was modeled using a combination of Shirley- and Tougaard-type shapes. In all the cases, prior to measurements, the samples was cleaved *in situ* under UHV at pressures below $1 \times 10^{-8}$ Pa.

The electronic structure supercell calculations with impurities were performed using the OpenMX code which provides a fully relativistic DFT implementation with localized pseudoatomic orbitals [42–44] and norm-conserving pseudopotentials[45]. The exchange correlation energy in the PBE version of generalized gradient approximation (GGA) was employed.[46] Structural parameters from Ref. [2] with modified vdW gap between SLs (2.89 Å) were used for $MnBi_2Te_4$ unit cell. For impurity calculations 2 x 2 supercells of $MnBi_2Te_4$, which provide 4 non-equivalent positions of Mn atoms in each layer, were used. The Ge concentrations of 25, 50 and 75% were obtained by replacing one, two and three Mn atoms with Ge atoms, respectively. The accuracy of the real-space numerical integration was specified by the cutoff energy of 450 Ry, the total energy convergence criterion was $10^{-6}$ eV. The k-mesh for Brillouin zones were specified as follows: 5 x 5 x 5 mesh for pristine $MnBi_2Te_4$ slab and 3 x 3 x 1 mesh for 2 x 2 supercells. The basis functions were taken as Bi8.0—s3p2d2f1, Te7.0—s3p2d2f1, Mn6.0—s3p2d1, Ge7.0—s3p2d2 (the pseudopotential cutoff radius is followed by a basis set specification). The Mn 3d states were treated within the DFT + U approach [47] within the Dudarev scheme [48] where U parameter equals 5.4 eV [2].

The band structure of $(Mn_{1-x}Ge_x)Bi_2Te_4$ with fully disordered impurities from first-principles calculations was obtained utilizing the spin polarized relativistic Korringa–Kohn–Rostoker (SPR-KKR) code, version 7.7. [49] The GGA in the PBE parametrization was used [46] with U correction and the atomic-sphere approximation (ASA) in the fully-relativistic approach. The coherent potential approximation (CPA) [50] was used to simulate chemical disorder. We used the basic functions up to $l$ = 3, a regular k-point grid with 250 points, and 30 energy points.

For the topological classification of $GeBi_2Te_4$ and $MnBi_2Te_4$, the electronic structure was calculated using Quantum Espresso code.[51,52] The tight-binding Hamiltonian was obtained using Wannier90 code.[53] Calculations of $Z_2$ invariants were done using WannierTools package.[54]

Acknowledgments

ARPES studies was partially supported by the Ministry of Science and Higher Education of the Russian Federation (No. FSMG-2023-0014). The Scanning Tunneling Spectroscopy at 1K supported by RSF (No. 23-72-30004 https://rscf.ru/en/project/23-72-30004/). Theoretical calculations supported by Saint Petersburg State University (Grant No. 94031444).

We thank Helmholtz-Zentrum Berlin and ELETTRA for granting access to the beamlines UE112-PGM2a (1^3), RGBL and BaDElPh and for provision of beamtimes. We are appreciated to Dr. Anna Makarova, Dr. Vera Neudachina and Elena Ushakova for the participation in spectra measurements. Calculations were performed using facilities of the Joint Supercomputer Center of Russian Academy of Sciences.